\documentclass[twocolumn,showpacs,showkeys,prd,aps,amsmath,amssymb,10pt]{revtex4-1}
\usepackage{graphicx}
\usepackage{color}
\usepackage{multirow}
\usepackage{bm}
\usepackage{mathtools}
\usepackage{subfig}
\begin{document}
\title{Analysis of mechanisms that could contribute to neutrinoless double-beta decay}
\author{Mihai Horoi}
\email{mihai.horoi@cmich.edu}
\author{Andrei Neacsu}
\email{neacs1a@cmich.edu}
\affiliation{Department of Physics, Central Michigan University, Mount Pleasant, Michigan 48859, USA}
\date{\today}
\begin{abstract}
Neutrinoless double-beta decay is a beyond the Standard Model process that would indicate that neutrinos are Majorana fermions, and the lepton number is not conserved. 
It could be interesting to use the neutrinoless double-beta decay observations to distinguish between several beyond Standard Model 
mechanisms that could contribute to this process. Accurate nuclear structure calculations of the nuclear matrix elements necessary to analyze the decay 
rates could be helpful to narrow down the list of contributing mechanisms. We investigate the information one can get from the angular and energy distribution of 
the emitted electrons and from the half-lives of several isotopes, assuming that the right-handed currents exist. For the analysis of these distributions, 
we calculate the necessary nuclear matrix elements using shell model techniques, and we explicitly consider interference terms.
\end{abstract}
\maketitle

\renewcommand*\arraystretch{1.3}

\section{Introduction}

Neutrinoless double-beta decay, if observed, would signal physics beyond the Standard Model (SM) that could be discovered at energies significantly lower than 
those at which the relevant degrees of freedom could be excited. The black-box theorems \cite{SchechterValle1982,Nieves1984,Takasugi1984,Hirsch2006} would indicate 
that the neutrinos are Majorana fermions, and the lepton number is violated in this process by two units.

However, it could be challenging to further use the neutrinoless double-beta decay observations to distinguish between many beyond Standard Model mechanisms that 
could contribute to this process \cite{Vergados2012,Horoi2013}. Accurate nuclear structure calculations of the nuclear matrix elements (NME) necessary to analyze the 
decay rates could be helpful to narrow down the list of contributing mechanisms and to better identify the more exotic properties of the neutrinos, such as the 
existence of the heavy sterile partners that could interact through right-handed currents \cite{KeungSenjanovic1983,Barry2013,Deppisch2016}. The NME for the standard 
mass mechanism were thoroughly investigated using several nuclear structure models. Figure 13 of Ref. \cite{NeacsuHoroi2015}  shows some of these NME for isotopes of 
immediate experimental relevance. Here, we describe the status of the  shell model calculations of these NME 
\cite{Horoi2013,NeacsuHoroi2015,BrownHoroiSenkov2014,SenkovHoroi2014,SenkovHoroiBrown2014,SenkovHoroi2013,HoroiBrown2013,NeacsuStoicaHoroi2012,HoroiStoica2010,HoroiStoicaBrown2007} 
and their relevance for discriminating possible competing mechanisms that may contribute to the neutrinoless double-beta decay process. 

One possible alternative/competing mechanism considers the contribution from the exchange of the heavy, mostly sterile, 
neutrinos \cite{KeungSenjanovic1983,Barry2013,Deppisch2016}. The exchange of left-handed heavy neutrinos is shown to be negligible in most 
cases \cite{Mitra2012,Blennow2010}. The exchange of the right-handed heavy neutrinos is predicted by left-right symmetric 
models \cite{PatiSalam1974,MohapatraPati1975,Senjanovic1975,KeungSenjanovic1983,Barry2013}, which are presently under active investigation at 
LHC \cite{CMS2014,Deppisch2016}. In either case, the same heavy neutrino-exchange NME are necessary for the analysis of the data. For example, considering only the 
competition between the light left-handed neutrino-exchange mechanism and the heavy right-handed neutrino-exchange mechanism, one could identify the dominant effect 
using half-lives of several isotopes, such as $^{76}$Ge and $^{136}$Xe \cite{Faessler2011}. Some of these heavy neutrino-exchange NME for isotopes of immediate 
experimental relevance are shown in Fig. 14 of Ref. \cite{NeacsuHoroi2015}. The range of these matrix elements is quite large due to their sensibility to the 
short-range correlation effects that were not treated consistently. One important improvement of these calculations would be obtaining an effective transition 
operator that takes into account consistently the short-range correlation effects and the effects of the missing single particle orbits from the model space \cite{HoltEngel2013}. 

Some other low-energy effects of the left-right symmetric models, such as those due to the so called $\lambda$ and $\eta$ mechanisms \cite{Doi1985,Barry2013}, 
could be identified experimentally if one could measure the angular and the energy distribution of the emitted electrons \cite{snemo2010}, but the analysis requires knowledge of additional NME that one can calculate. 
Finally, some more exotic possibilities \cite{Vergados2012,Prezeau2003} leading to one- and two-pion exchange NME \cite{Gehman2007} were also calculated in the past within 
the interacting shell model approach \cite{Horoi2013,HoroiBrown2013}, and quasiparticle random phase approximation (QRPA) (see, e.g., Ref. \cite{Vergados2012} and references 
therein). A more general approach that includes a complete set of dimension six and dimension nine operators to the SM Lagrangian, as well as R-parity violating SUSY contributions, Kaluza-Klein modes in higher dimensions \cite{Bhattacharyya2003,Deppisch2007}, violation of Lorentz invariance and equivalence principle \cite{Leung2000,Klapdor1999,Barenboim2002}, is given in Refs. \cite{Pas1999,Deppisch2012}. Information from double-beta decay can help constrain these contributions, but additional information from the colliders is needed for a full analysis.

In this paper, we consider the possibility of disentangling the contributions of the right-handed currents to the neutrinoless double-beta decay process. 
Our analysis mostly focuses on the information one can get from the two-electron energy and angular distributions, which could be used to distinguish contributions 
coming from the $\lambda$ and $\eta$ mechanisms from those of the usual light neutrino-exchange mechanism. The analysis is done for $^{82}$Se, which was chosen as a 
baseline isotope by the SuperNEMO experiment \cite{snemo2010,nemo32014}. During the preparation of this manuscript, we also found a more general analysis of the terms 
contributing to the angular and energy distributions for most of the double-beta decay isotopes based on improved phase space factors and QRPA NME \cite{Stefanik2015}. 
Efforts of separating these effects are not new (see, e.g., Refs. \cite{Doi1983,TOMODA1986,Hirsch1994,PhysRevC.64.035501,Bilenky2004} among others). 
Our analysis is however more detailed and more specific to the decay of the $^{82}$Se isotope. It considers the competitions between the mass mechanisms and the 
heavy right-handed neutrino-exchange mechanism if the contributions from $\lambda$ and $\eta$ mechanisms are ruled out by the two-electron angular and energy distributions. 

The paper is organized as follows: Section II presents the general formalism used to describe the neutrinoless double-beta decay under the assumption that the right-handed 
currents would contribute. Section III describes the associated two-electron angular and energy distributions. Section IV analyzes the two-electron angular and energy 
distributions for different scenarios that consider different relative magnitudes of the $\lambda$ and $\eta$ mechanism amplitudes (please notice the changes of notation). 
Section V considers the possibility of disentangling the mass mechanisms from the heavy right-handed neutrino-exchange mechanism, if the $\lambda$ and $\eta$ contributions 
could be ruled out by the two-electron energy and angular distributions. Section VI is devoted to conclusions, and Appendixes A, B, and C present detailed formulas used in the formalism.

\section{$0\nu\beta\beta$ decay formalism}\label{formalism}

If right-handed currents exist there are several possible contributions to the neutrinoless double-beta decay rate \cite{Doi1983,Doi1985}. Usually, only the light left-handed neutrino-exchange mechanism (a.k.a. the mass mechanism) is taken 
into consideration, but other mechanisms could play a significant role \cite{Vergados2012}. One popular model that considers the right-handed currents contributions is the left-right symmetric model \cite{MohapatraPati1975,Senjanovic1975}, which assumes the existence of heavy particles that are not part of the Standard Model (see also Ref. \cite{Barry2013} for a review specific to double-beta decay).

In the framework of the left-right symmetric model one can write the electron neutrino fields (see Appendix A where we use the notations of Ref. \cite{Barry2013}) as

\begin{eqnarray}
\label{nfields}
\nonumber \nu\sp{\prime}_{eL}&=&\sum_{k}^{light} U_{ek}\nu_{kL}+\sum_k^{heavy}S_{ek} N^c_{kR},\\
\nu\sp{\prime}_{eR}&=&\sum_{k}^{light} T^*_{ek}\nu^c_{iL}+\sum_k^{heavy}V^*_{ek} N_{kR},
\end{eqnarray}
where $\nu\sp{\prime}$ represent flavor states, $\nu$ and $N$ represent mass eigenstates, $U$ and $V$ mixing matrices are almost unitary, while $S$ and $T$ mixing matrices are small. 
The $\nu\sp{\prime}_{eL}$ electron neutrino is active for the $V-A$ weak interaction and sterile for the $V+A$ interaction, with the opposite being true for $\nu\sp{\prime}_{eR}$. 
Then, the neutrinoless half-life  expression is given by

\begin{eqnarray}
\label{flifetime}
\nonumber \left[ T^{0\nu}_{1/2} \right] ^{-1}  =  G^{0\nu}_{01} g^4_A & \mid & M^{0\nu}\eta_{\nu} +M^{0N}\left(\eta^L_{N_R}+\eta^R_{N_R}\right) \\
 & & +\eta_{\lambda}X_{\lambda}+\eta_{\eta}X_{\eta}+\cdots \mid ^2  ,
\end{eqnarray}
where $\eta_{\nu}$, $\eta^L_{N_R}$, $\eta^R_{N_R}$, $\eta_{\lambda}$, and $\eta_{\eta}$ are neutrino physics parameters defined in Ref. \cite{Barry2013}. 
See Appendix A for the definition of the neutrino physics parameters. One should mention that our $\eta_{\lambda}$ and $\eta_{\eta}$ parameters correspond to $\lambda$ 
and $\eta$ of Ref. \cite{snemo2010}. Above, $M^{0\nu}$  and $M^{0N}$ are the light and heavy neutrino-exchange nuclear matrix elements \cite{Horoi2013,SenkovHoroiBrown2014,Vergados2012} (see their explicit decomposition in Appendix B), and $X_{\lambda}$ and $X_{\eta}$ represent combinations of NME and phase space factors that are analyzed below. Here, $G_{01}^{0\nu}$ is a phase space factor \cite{SuhonenCivitarese1998} that can be calculated with relatively good precision in most cases \cite{Kotila2012,StoicaMirea2013}, and $g_A=1.27$ (see also Appendix C). The ``$\cdots $'' sign stands for other possible contributions, such as those of R-parity violating SUSY particle exchange \cite{Vergados2012,Horoi2013}, Kaluza-Klein modes \cite{Bhattacharyya2003,Deppisch2007,Horoi2013}, violation of Lorentz invariance, equivalence principle \cite{Leung2000,Klapdor1999,Barenboim2002},etc, which are neglected here. 

The $\eta^L_{N_R}$ term also exists in the seesaw type I mechanisms, but its contribution is negligible if the heavy mass eigenstates are larger than 1 GeV \cite{Blennow2010}. Assuming a seesaw type I dominance \cite{DevMitra2015}, we neglect it here. If the $\eta_{\lambda}$ and $\eta_{\eta}$ contributions could be ruled out by the two-electron energy and angular distributions, the remaining $\eta_{\nu}$ and $\eta^R_{N_R}$ terms have a very small interference contribution (the interference term is at most 8\% of the two terms in the parenthesis of Eq. \ref{hlifetime} \cite{Halprin1983,Faessler2011}), and the half-life becomes

\begin{flalign} \label{hlifetime}
\left[ T^{0\nu}_{1/2} \right] ^{-1}  =  G^{0\nu}_{01} g^4_A  \left( \left| M^{0\nu}\right|^2 \left| \eta_{\nu} \right|^2 + \left| M^{0N} \right|^2 \left| \eta^R_{N_R}\right|^2 \right). & & 
\end{flalign}

Then, the relative contribution of the $\eta_{\nu}$ and $\eta^R_{N_R}$ can be gauged out if one measures the half-life of at least two isotopes \cite{Faessler2011,Vergados2012}, provided that the corresponding matrix elements $M^{0\nu}$ and $M^{0N}$ are known with good precision (see Sec. \ref{heavy} below). These matrix elements were calculated using several methods including the interacting shell model (ISM) \cite{Blennow2010,Horoi2013,HoroiBrown2013,SenkovHoroiBrown2014,SenkovHoroi2013,BrownHoroiSenkov2014,NeacsuHoroi2015} (see Ref. \cite{NeacsuHoroi2015} for a review), quasiparticle random phase approximation (QRPA) \cite{Vergados2012,Suhonen2015}, and interacting boson model (IBM) \cite{Barea2015}. In general, the ISM  results for $M^{0\nu}$ are quite close one to another but smaller than the QRPA and IBM results; the ISM and IBM results for $M^{0N}$ are close, while they are both smaller than the QRPA results. An explanation of this behavior was recently provided \cite{BrownFangHoroi2015}, which suggests a path for improving  these NME. We believe that nuclear shell model matrix elements are the most reliable because they take into consideration all correlations around the Fermi surface, respect all symmetries, and take into account consistently the effects of the missing single particle space via many-body perturbation theory (shown to be small, about 20\%,  for $^{82}$Se \cite{HoltEngel2013}). Because of that, we use no quenching for the bare $0\nu\beta\beta$ operator in our calculations. This conclusion is different from that for the simple Gamow-Teller operator used in single beta and $2\nu\beta\beta$ decays for which a quenching factor of about 0.7 is necessary \cite{BrownFangHoroi2015}.

In what follows, we provide an analysis of the two-electron relative energy and angular distributions using shell model NME. This analysis could be used to analyze data that may be provided by the SuperNEMO experiment to identify the relative contributions of $\eta_{\lambda}$ and $\eta_{\eta}$ terms in Eq. (\ref{flifetime}). A similar analysis using QRPA NME was given in Ref. \cite{snemo2010}. During the preparation of this manuscript, we also found a more general analysis of the terms contributing to the angular and energy distributions, for most of the double-beta decay isotopes, based on improved phase space factors and QRPA NME \cite{Stefanik2015}. However, our analysis is more detailed and more specific to the decay of the $^{82}$Se isotope. The starting point is provided by the classic paper of Doi, Kotani, and Tagasuki \cite{Doi1985}, which describes the neutrinoless double-beta decay process using a low-energy Hamiltonian that includes the effects of the right-handed currents. The $\eta_{\lambda}$ and $\eta_{\eta}$ terms in Eq. (\ref{flifetime}) are related to the $\lambda$ and $\eta$ terms in Ref. \cite{Doi1985}. With some simplifying notations the half-life expression \cite{Doi1985} (here, we omit the contribution from the $\eta^L_{N_R}$ term, which has the same energy and angular distribution as the $\eta_{\nu}$ term) is given by

\begin{eqnarray}
\label{lifetime}
\nonumber \left[ T^{0\nu}_{1/2} \right] ^{-1}&=&\left|M_{GT}^{0\nu}\right| ^2 \left\{ C_{\nu^2} + C_{\nu\lambda}\textmd{cos}\phi_1 + C_{\nu\eta}\textmd{cos}\phi_2 \right.  \\
  &+& \left. C_{\lambda^2} + C_{\eta^2} +C_{\lambda\eta}\textmd{cos}(\phi_1 - \phi_2) \right\} ,
\end{eqnarray}
where $\phi_1$ and $\phi_2$ are the relative $CP$-violating phases (\ref{phases}), and $M_{GT}^{0\nu}$ is the Gamow-Teller contribution of the light neutrino-exchange NME.
Different processes give rise to several contributions: $C_{\nu^2}$ are from the left-handed leptonic and currents, $C_{\lambda^2}$ from the right-handed leptonic and right-handed hadronic currents, and $C_{\eta^2}$ from the right-handed
 leptonic and left-handed hadronic currents. Interference between these terms is represented by the the contributions of
 $C_{\nu\lambda}$, $C_{\nu\eta}$ and $C_{\lambda\eta}$. The precise definitions are
\begin{eqnarray} \label{c-mechanism}
\nonumber C_{\nu^2}	&= C_1 \left< \nu\right> ^2, C_{\nu\lambda}	=& C_2\left< \nu\right>\left< \lambda\right>, 	C_{\nu\eta}	= C_3\left< \eta \right>\left< \nu\right>, \\
      C_{\lambda^2}	&= C_4\left< \lambda\right>^2, C_{\eta^2}	=& C_5\left< \eta \right>^2, 	\ \ \		C_{\lambda\eta}	= C_6\left< \eta \right>\left< \lambda\right> ,
\end{eqnarray}
where $C_{1-6}$ are combinations of nuclear matrix elements and phase-space factors (PSF).
Their expressions can be found in Appendix B, Eqs. (\ref{ci-uri}). Here, $M_{GT}^{0\nu}$ and the 
other nuclear matrix elements that appear in the expressions of the $C$ factors are presented in Eq. (\ref{nme-gt}). In the context of the left-right symmetric model, we associate the neutrino physics parameters $\left< \nu \right>$, $\left< \lambda \right>$, and $\left< \eta \right>$ with the corresponding $\eta_i$ parameters defined in Appendix A,
\begin{subequations}\label{mass-parameters}
\begin{align}
\left< \nu \right>	& = \left| \eta_{\nu} \right|\ ,\\\
\left< \lambda \right>	& = \left| \eta_{\lambda} \right|\ ,\\
\left< \eta \right>	& = \left| \eta_{\eta} \right|\ ,
\end{align}
\end{subequations}
but we leave them in this generic form for the case that other mechanisms could contribute. For example, any contribution from a mechanism whose amplitude is proportional with $\sqrt{G^{0\nu}_{01}}$, such as $\eta^L_{N_R}$ and $\eta^R_{N_R}$, may be added to the $\left< \nu \right>$ term with an appropriate redefinition of the nuclear matrix elements and the interference phases.

 
\section{$0\nu\beta\beta$ decay electrons distributions}

The differential decay rate of the $0^+\rightarrow0^+$ $0\nu\beta\beta$ transition can be expressed as:
\begin{equation}\label{diff-rate}
\frac{\textmd{d}^2W_{0^+\rightarrow0^+}^{0\nu}}{\textmd{d}\epsilon_1 \textmd{d}\text {cos}\theta_{12}} = \frac{a_{0\nu} \omega_{0\nu}(\epsilon_1)}{2\left( m_eR \right)^2} \left[A(\epsilon_1)+B(\epsilon_1)\text {cos}\theta_{12} \right]  .
\end{equation}
Here, $\epsilon_1$ is the energy of one electron in units of $m_e c^2$, $R$ is the nuclear radius ($R=r_0 A^{1/3}$, with $r_0=1.2$fm), $\theta_{12}$ is the angle between the outgoing electrons, and the expressions for the constant $a_{0\nu}$ and the function $\omega_{0\nu}$ are 
given in the Appendix C, Eqs. (\ref{a0}) and (\ref{omega0}), respectively.
The functions $A(\epsilon)$ and $B(\epsilon)$ are defined as combinations of factors that include PSF and NME:
\begin{subequations}\label{ab}
\begin{align}
 A(\epsilon_1)&=|N_1(\epsilon_1)|^2+|N_2(\epsilon_1)|^2+|N_3(\epsilon_1)|^2+|N_4(\epsilon_1)|^2,  \\
 B(\epsilon_1)&=-2\text{Re}\left[ N_1^\star(\epsilon_1) N_2(\epsilon_1)+N_3^\star(\epsilon_1) N_4(\epsilon_1) \right]. 
\end{align}
 \end{subequations}
The detailed expressions of the $N_{1-4}(\epsilon_1)$ components are presented in Eqs. (\ref{n-uri}).

The expression of the half-life can be written as follows:
\begin{eqnarray} \label{half-life}
\nonumber \left[ T_{1/2}^{0\nu} \right]^{-1} &=& \frac{1}{\text{ln2}}\int \textmd{d}W_{0^+\rightarrow0^+}^{0\nu} =  \frac{a_{0\nu}}{\text{ln2} \left( m_eR \right)^2}\\
&\times& \int_1^{T+1} A(\epsilon_1) \omega_{0\nu}(\epsilon_1)\textmd{d}\epsilon_1,
\end{eqnarray} 
with the kinetic energy T defined as
\begin{equation}\label{t-energy}
 T=\frac{Q_{\beta\beta}}{m_e c^2}.
 \end{equation} 

 \subsection{Angular distributions}

The integration of Eq. (\ref{diff-rate}) over $\epsilon_1$ provides the angular distribution of the electrons. 
We can now write it as
\begin{eqnarray}
 \nonumber \frac{\textmd{d}W_{0^+\rightarrow0^+}^{0\nu}}{\textmd{d}\Omega}&=&\frac{a_{0\nu}}{4\pi \left( m_eR \right)^2}\left[ 
 \int_1^{T+1} A(\epsilon_1)\omega_{0\nu}(\epsilon_1)\textmd{d}\epsilon_1 \right. \\ 
 &+& \left. \frac{\textmd{d}\Omega}{2\pi} \int_1^{T+1} B(\epsilon_1)\omega_{0\nu}(\epsilon_1)\textmd{d}\epsilon_1 \right],
\end{eqnarray}
where $d\Omega=2\pi d\text {cos}\theta_{12}$.
 
\subsection{Energy distributions} 
Integrating Eq. (\ref{diff-rate}) over cos$\theta_{12}$, one obtains the single electron spectrum.
When investigating the energy distribution, it is convenient to express the decay rate as a function of 
the difference in the energy of the two outgoing electrons, $\Delta t=(\epsilon_1 - \epsilon_2 )m_e c^2$, where
$ \epsilon_2=T+2-\epsilon_1$ is the kinetic energy of the second electron.
We now express the energy of one electron as
\begin{equation}
\epsilon_1=\frac{T+2+\frac{\Delta t}{m_e c^2}}{2}.
\end{equation}
After changing the variable, the energy distribution as a function of $\Delta t$ is
\begin{equation}
 \frac{2\textmd{d}W_{0^+\rightarrow0^+}^{0\nu}}{\textmd{d}(\Delta t)}=\frac{2a_{0\nu}}{\left( m_eR \right)^2}
 \frac{\omega_{0\nu}(\Delta t)}{m_e c^2} A(\Delta t).
\end{equation}

\begin{figure}[htb]
\includegraphics[width=0.98\linewidth]{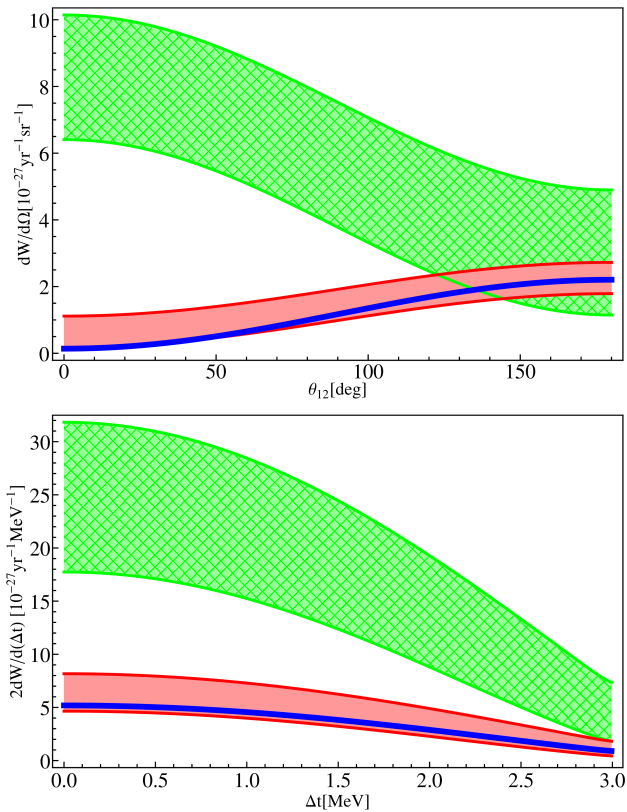}
\caption{Electrons angular distribution (upper panel) and energy distributions (lower panel) for the competition between $\nu$ and $\eta$ mechanisms, Case 1 (see Sec. IV for a full description of the bands).}
\label{eta} 
\end{figure}

\begin{figure}[htb]
\includegraphics[width=0.98\linewidth]{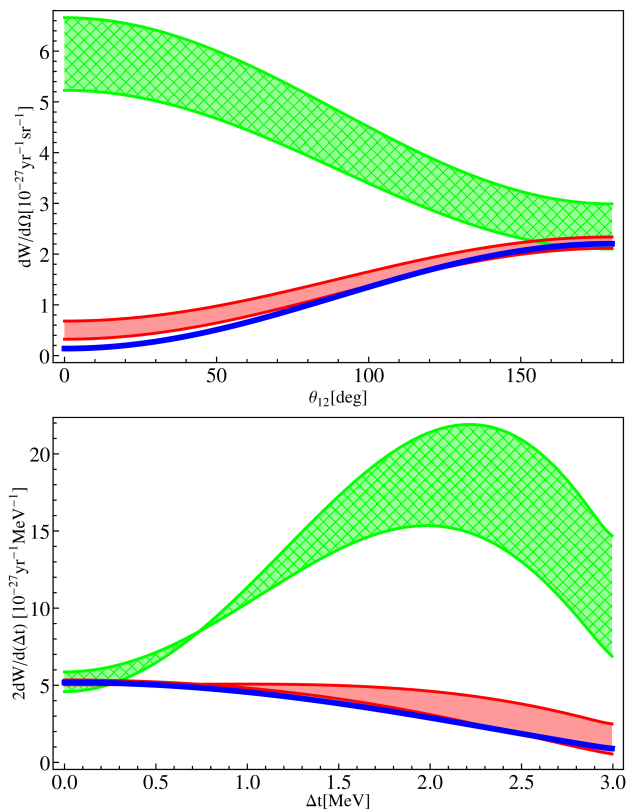}
\caption{Same as Fig. \ref{eta} for the competition between $\nu$ and $\lambda$ mechanisms, Case 2.}\label{lambda}
\end{figure}

\section{Results}
Here, we analyze in detail the two-electron angular and energy distributions for $^{82}$Se, which was chosen as a baseline isotope by the SuperNEMO experiment \cite{snemo2010,nemo32014}. 
We calculate the $^{82}$Se NME of Eq. (\ref{nme-gt}) using a shell model approach with the JUN45 \cite{JUN45} effective Hamiltonian 
in the $jj44$ model space \cite{SenkovHoroiBrown2014,SenkovHoroi2014}. The nuclear structure effects are taken into account by the inclusion of short-range correlations with CD-Bonn 
parametrization, finite nucleon size effects, and higher order corrections of the nucleon current \cite{HoroiStoica2010}. 
We point out that some of the neutrino potentials in Eq. (\ref{potentials}) are divergent \cite{Doi1985}, such that
the approximations $\chi_{GT\omega}=2-\chi_{GTq}$ and $\chi_{F\omega}=2\chi_F-\chi_{Fq}$ \cite{Tomoda1991} are not accurate. 
This simplification was widely used because of the high complexity and difficulty of the previous shell model calculations with large model spaces \cite{Retamosa1995,Caurier1996} when most of $0\nu\beta\beta$ decaying isotopes were considered. 
A solution to this problem is to first perform the radial integral over the coordinate space and only after the second integral over the momentum space in Eq. (\ref{i-alpha}).
For $g_A$ we use the older value of 1.254 for an easier comparison to other NME and PSF results in the literature. It was shown in Ref. 
\cite{SenkovHoroiBrown2014} that changing to the newer value of 1.27 \cite{pdg-2014} changes the result by only 0.5\%. Most of uncertainties in the shell model calculations come from different parametrization of the short range correlations, but they are less than 20\% for most of the NME. It is also worth noting that the shell model NME are in general smaller by a factor of 2 than the QRPA NME, but recent work on restoring the broken symmetries in QRPA shows a tendency of reducing the QRPA values towards the shell model ones (see, e.g., Sec. IV.c of Ref. \cite{HoroiNeacsu2016}).
 
The NME calculated in this work are presented on the first line of Table \ref{tab-nme}. The second line displays the normalized values $\chi_{\alpha}$ ($\alpha=F,GT\omega,F\omega,GTq,Fq,T,R,P$).
\begin{table}[htb] 
 \caption{The $^{82}$Se NME corresponding to Eq. (\ref{chi-form}).}
 \begin{tabular}{ccccccccc} \hline \hline \label{tab-nme}
  $M_{GT}$&$M_{F}$&$M_{GT\omega}$&$M_{F\omega}$&$M_{GTq}$&$M_{Fq}$&$M_{T}$&$M_{R}$&$M_{P}$ \\ \hline 
   $2.993$&$-0.633$& $2.835$        & $-0.618$      & $3.004$   & $-0.487$ &$0.012$  &$3.252$  &$-1.286$   \\ \hline
	  &$\chi_{F}$&$\chi_{GT\omega}$&$\chi_{F\omega}$&$\chi_{GTq}$&$\chi_{Fq}$&$\chi_{T}$&$\chi_{R}$&$\chi_{P}$ \\ \hline 
  	  &$-0.134$    & $0.947$	       &$-0.131$		&$1.003$	     &$-0.103$	 & $0.004$    &$1.086$     & $0.430$ \\ \hline \hline
 \end{tabular} \\
The values of the $\chi_{1\pm}$ and $\chi_{2\pm}$ factors of Eq. (\ref{chi}) are \\
$\chi_{1+}=0.717$, $\chi_{1-}=1.338$, $\chi_{2+}=0.736$, $\chi_{2-}=0.930$. \\ 
\end{table}

The PSF that enter in the components of Eq. (\ref{lifetime}) are calculated in this work using Eq. (\ref{psfs-form}).
These can be also calculated by a simple manipulation of Eq. (\ref{half-life}), involving $\tilde{A}_{\pm k}$ defined in Appendix B.
Using a new effective method to calculate PSF \cite{HoroiNeacsuPSF2015} in agreement with other recent results, we choose a value of 92 for the effective ``screening factor'' ($S_f$) that changes the charge of the daugther nucleus, $Z_s=\frac{S_f}{100} Z$. Reference \cite{HoroiNeacsuPSF2015} provides a detailed study of the $2\nu\beta\beta$ and $0\nu\beta\beta$ PSF using this method for 11 nuclei.
In the case of $G_1$, we obtain results  which are in good agreement with those of Ref. \cite{Stefanik2015}, having a difference of about 8\%. 
The results of Ref. \cite{Stefanik2015} have been obtained more rigorously by solving numerically the Dirac equation and by including the effects of the finite nuclear size and electron screening using a Coulomb potential derived from a realistic proton density distribution in the daughter nucleus. 
The largest difference is 15.5\% in the case of $G_8$. The original formalism of Ref. \cite{Doi1985} provides significantly larger differences, of up 
to more than 64\% for $G_8$ of $^{82}$Se, and would result in differences in half-lives of over 30\% for Case 4, where all the nine PSF contribute.
However, given the larger uncertainty in the NME \cite{BrownFangHoroi2015}, our approximation is satisfactory and we use it in calculations of the half-lives and of the two-electron angular and energy distributions.

\begin{figure}[htb] 
\includegraphics[width=0.98\linewidth]{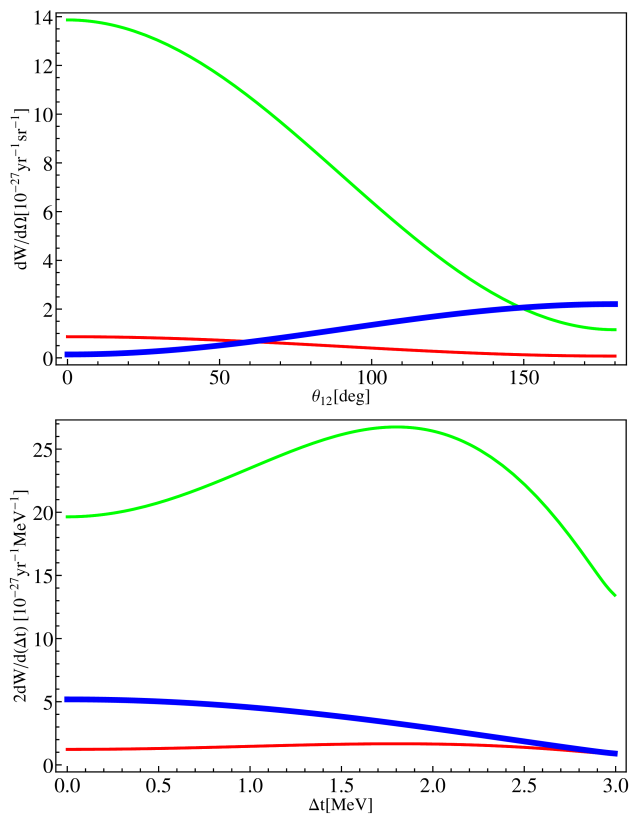}
\caption{Same as Fig. \ref{eta} for the competition between $\lambda$ and $\eta$ mechanisms, Case 3.}\label{no_mass}
\end{figure}

\begin{figure}[htb]
\includegraphics[width=0.97\linewidth]{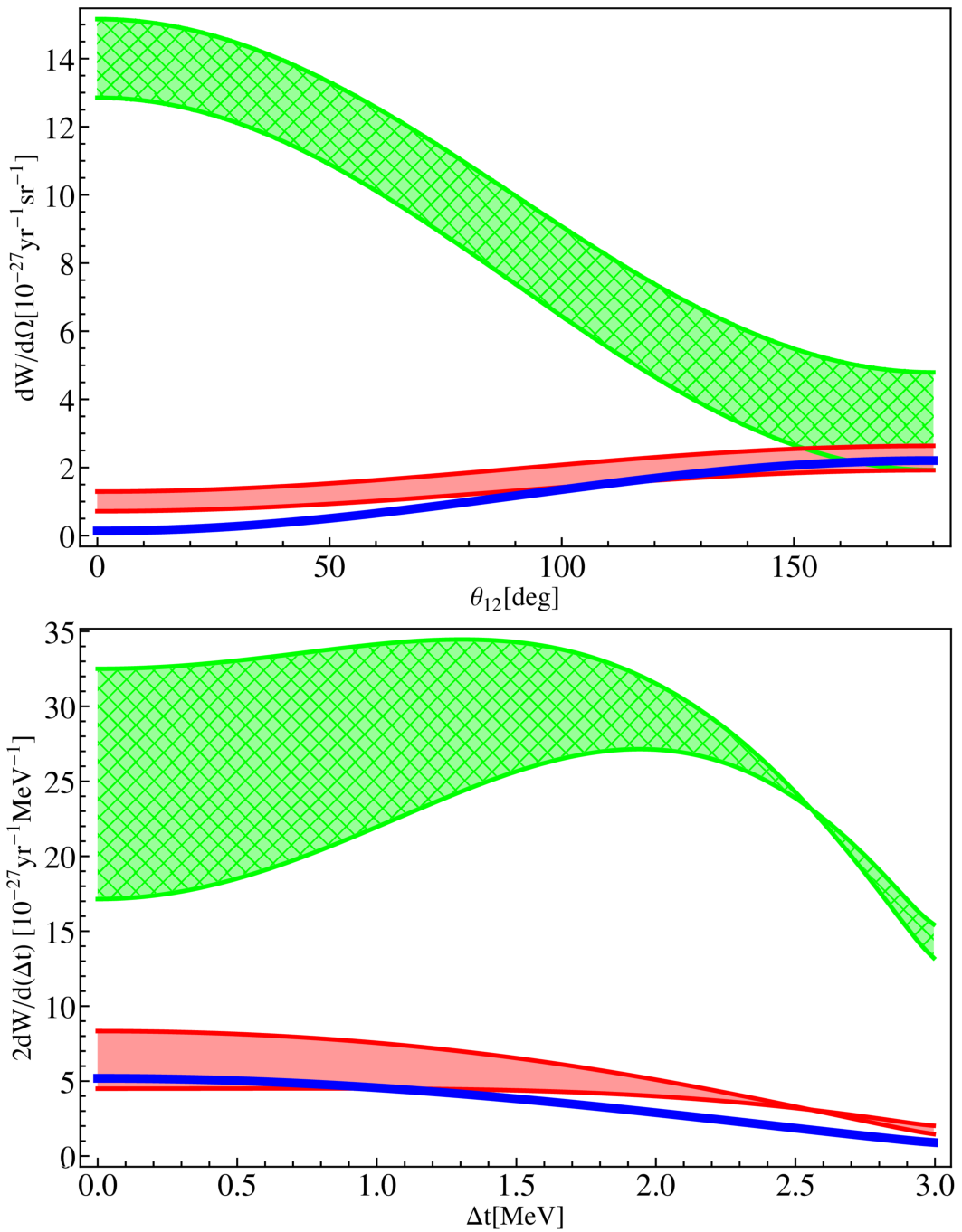}
\caption{Same as Fig. \ref{eta} for the competition between $\nu$, $\lambda$ and $\eta$ mechanisms, Case 4.}\label{all_mech}
\end{figure}

\begin{table}[htb]
  \caption{The $^{82}$Se PSF corresponding to Eq.  (\ref{psfs-form}) expressed in $\left[ \textmd{yr}^{-1}\right] $}
  \begin{tabular}{@{}ccccc@{}} \hline \hline \label{tab-psfs}
$G_1\times 10^{14}$	&$G_2\times 10^{14}$	&$G_3\times 10^{14}$	&$G_4\times 10^{15}$	&$G_5\times 10^{13}$ \\ \hline
$2.31$			&$7.93$			&$1.61$			&$4.75$			&$5.33$ \\ \hline
			&$G_6\times 10^{12}$	&$G_7\times 10^{10}$	&$G_8\times 10^{11}$	&$G_9\times 10^{9}$ \\ \hline 
			&$4.09$			&$2.97$			&$2.02$			&$1.09$  \\ \hline \hline	  
 \end{tabular}
\end{table}

In our analysis of the angular and energy distributions we consider five scenarios:
a reference case named ``Case 0'', commonly referred to in the literature as the ``mass mechanism''(displayed with a thick blue line in all the figures);
a case when only the mass mechanism and the $\eta$ mechanism contribute, presented as ``Case 1''; 
the scenario when only the mass mechanism and the $\lambda$ mechanism contribute, ``Case 2``;
the case when the mass mechanism does not contribute and we have competition and interference between the $\lambda$ and the $\eta$ mechanisms denoted as ''Case 3``;
and the most complex scenario, ''Case 4``, when there is competition and interference between all the mechanisms. 

\begin{table}[htb]
 \caption{The neutrino parameter values chosen for the five cases described in the text.}
 \begin{tabular}{@{}lccc@{}} \hline \hline \label{cases}
		&$\left< \nu\right>$&$\left< \lambda\right>$& $\left< \eta\right>$\\ \hline
Case 0 Blue	&$2\times 10^{-7}$&$0$			&$0$			\\ \hline
Case 1 Red	&$2\times 10^{-7}$&$0$			&$0.5\times 10^{-9}$	\\
Case 1 Green	&$2\times 10^{-7}$&$0$			&$2\times 10^{-9}$	\\\hline
Case 2 Red	&$2\times 10^{-7}$&$0.5\times 10^{-7}$	&$0$			\\
Case 2 Green	&$2\times 10^{-7}$&$2\times 10^{-7}$	&$0$			\\ \hline
Case 3 Red	&$0$		&$0.5\times 10^{-7}$	&$0.5\times 10^{-9}$	\\
Case 3 Green	&$0$		&$2\times 10^{-7}$	&$2\times 10^{-9}$	\\ \hline
Case 4 Red	&$2\times 10^{-7}$&$0.5\times 10^{-7}$	&$0.5\times 10^{-9}$	\\
Case 4 Green	&$2\times 10^{-7}$&$2\times 10^{-7}$	&$2\times 10^{-9}$	\\ \hline \hline
\end{tabular}
\end{table}

The values of the effective parameters for these scenarios are chosen such that they highlight the competition or the dominance 
of these mechanisms, taking into account the current experimental limits \cite{Barry2013,Stefanik2015} for the $^{76}$Ge $0\nu\beta\beta$ half-life (see also Appendix A).
They are presented in Table \ref{cases}. In the figures, the red color indicates the lower values for $\lambda$ or $\eta$, while the green color is
used for the higher values. 

\begin{figure}[ht!]
\includegraphics[width=0.98\linewidth]{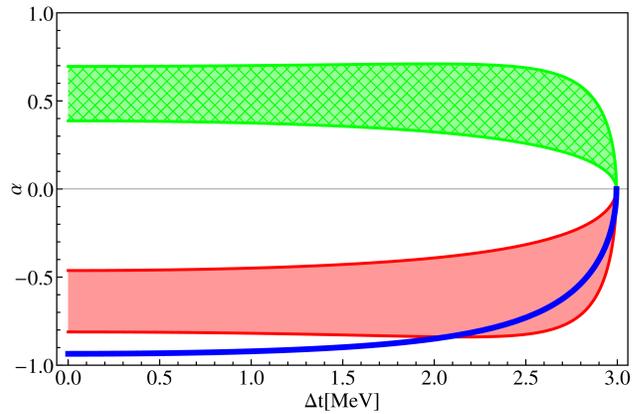}
\caption{The angular correlation coefficient corresponding to Case 1. The meaning of the bands is the same as in Figs. $1-4$.}
\end{figure}

\begin{figure}[ht!]
{\includegraphics[width=0.98\linewidth]{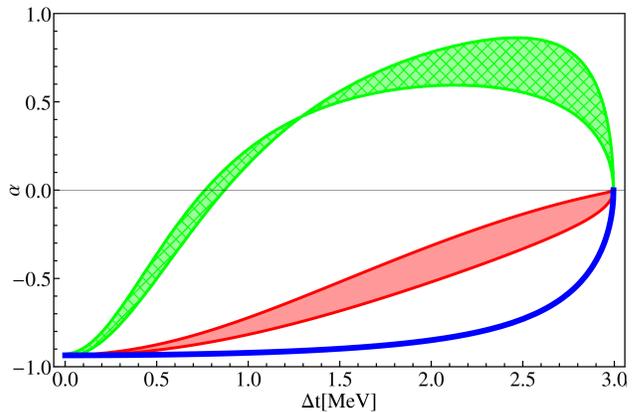}}
\caption{The angular correlation coefficient corresponding to Case 2. The meaning of the bands is the same as in Figs. 1-4.}
\end{figure}

For an easier evaluation of each contribution and the interference effects, we provide in Table \ref{tab-ci} the calculated $C_{i}$ factors ($i=1, \ldots , 6$)
of Eq. (\ref{ci-uri}), together with their effective values from Eqs. (\ref{c-mechanism}), for each particular case. 
Due to the large $G_7$,$G_8$, and $G_9$ PSF, the contribution of $C_5$ has a significantly higher magnitude compared to the other factors, such that the calculations are
very sensitive to the $\eta$ mechanism for the present limits of the neutrino physics parameters. 

\begin{table*}[htb]
 \caption{The $^{82}$Se $C_{i}$ factors ($i=1, \ldots , 6$) corresponding to Eq.  (\ref{lifetime}) expressed in $\left[ yr^{-1}\right] $. 
 We also present the effective values when these factors are multiplied with the neutrino parameters from the four cases discussed}
  \begin{tabular}{lcccccc} \hline \hline  \label{tab-ci}
	    &$C_1$		 &$C_2$			&$C_3$			&$C_4$			&$C_5$			&$C_6$ 			\\ \hline	       
	    &$2.94\times 10^{-14}$&$-1.46\times 10^{-14}$&$4.75\times 10^{-12}$	&$7.72\times 10^{-14}$	&$1.15\times 10^{-9} $	&$-1.01\times 10^{-13}$	\\ \hline
	    &$C_{\nu^2}$	 &$C_{\nu\lambda}$	&$C_{\nu\eta}$		&$C_{\lambda^2}$	&$C_{\eta^2}$		&$C_{\lambda \eta}$	\\ \hline
Case 0 Blue &$1.18\times 10^{-27}$&$0$			&$0$			&$0$			&$0$			&$0$			\\ \hline
Case 1 Red  &$1.18\times 10^{-27}$&$0$			&$4.75\times 10^{-28}$	&$0$			&$2.86\times 10^{-28}$	&$0$			\\
Case 1 Green&$1.18\times 10^{-27}$&$0$			&$1.90\times 10^{-27}$	&$0$			&$4.58\times 10^{-27}$	&$0$			\\ \hline
Case 2 Red  &$1.18\times 10^{-27}$&$-1.46\times 10^{-28}$&$0$			&$1.93\times 10^{-28}$	&$0$			&$0$			\\ 
Case 2 Green&$1.18\times 10^{-27}$&$-5.83\times 10^{-28}$&$0$			&$3.09\times 10^{-27}$	&$0$			&$0$			\\ \hline
Case 3 Red  &$0$		 &$0$			&$0$			&$1.93\times 10^{-28}$	&$2.86\times 10^{-28}$	&$-2.53\times 10^{-27}$	\\
Case 3 Green&$0$		 &$0$			&$0$			&$3.09\times 10^{-27}$	&$4.58\times 10^{-27}$	&$-4.04\times 10^{-29}$	\\ \hline 
Case 4 Red  &$1.18\times 10^{-27}$&$-1.46\times 10^{-28}$&$4.75\times 10^{-28}$	&$1.93\times 10^{-28}$	&$2.86\times 10^{-28}$	&$-2.53\times 10^{-30}$	\\
Case 4 Green&$1.18\times 10^{-27}$&$-5.83\times 10^{-28}$&$1.90\times 10^{-27}$	&$3.09\times 10^{-27}$	&$4.58\times 10^{-27}$	&$-4.04\times 10^{-29}$	\\ \hline \hline
\end{tabular} 
\end{table*}

One may calculate the $0\nu\beta\beta$ half-life with either Eq. (\ref{lifetime}) using the nine PSF of Eq. (\ref{psfs-form}) displayed in Table \ref{tab-psfs} or by integrating Eq. (\ref{diff-rate}) over angles ($\theta_{12}$ from $0$ to $\pi$) and energy  in Eq. (\ref{half-life}) ($\Delta t$ goes from $0$ to $Q_{\beta\beta}$, which is $2.99$ MeV for $^{82}$Se).
The calculated half-lives for the cases of interest are presented in Table \ref{half-lives}. There are four combinations for the $CP$ phases 
$\phi_1$ and $\phi_2$, providing up to four values for the half-lives for each case. 
All half-lives in Table \ref{half-lives}, except Case 3 Red, are above the present experimental limits, but within the reach of the SuperNEMO experimental setup ($1.0\times 10^{26}$ years).
One should also mention that the on-axis limits for the neutrino physics parameters $<\lambda>$ and $<\eta>$ corresponding to the same half-life, $9.41\times 10^{25}$ years, as the 100 meV mass mechanism are $1.2 \times 10^{-7}$ and $1.0 \times 10^{-9}$, respectively. 
The bands in the figures represent the interference 
effects of these phases, and their width is the maximum difference between them. 
In the case of the mass mechanism, there is no interference, such that Case 0 is represented by a single thick blue line. 
This line is present in all the figures to provide the reader with a reference scenario, which is the most studied in the literature.
In the following, we discuss these cases.

\begin{table*}[htb]
 \caption{Calculated half-lives ($T_{1/2}$) for the four possible combinations of values for $\phi_1$ and $\phi_2$ in Eq. (\ref{lifetime}).}
 \begin{tabular}{@{}l|c|c|c|c@{}} \hline  \hline \label{half-lives} 
		&$\phi_1=0,\phi_2=0$	&$\phi_1=\pi,\phi_2=\pi$&$\phi_1=0,\phi_2=\pi$	&$\phi_1=\pi,\phi_2=0$	\\ \hline
Case 0 Blue	&$9.41\times 10^{25}$	&$9.41\times 10^{25}$	&$9.41\times 10^{25}$	&$9.41\times 10^{25}$	\\ \hline
Case 1 Red	&$5.72\times 10^{25}$	&$1.12\times 10^{26}$	&$1.12\times 10^{26}$	&$5.72\times 10^{25}$	\\
Case 1 Green	&$1.45\times 10^{25}$	&$2.87\times 10^{25}$	&$2.87\times 10^{25}$	&$1.45\times 10^{25}$	\\\hline
Case 2 Red	&$9.05\times 10^{25}$	&$7.31\times 10^{25}$	&$9.05\times 10^{25}$	&$7.31\times 10^{25}$	\\
Case 2 Green	&$3.01\times 10^{25}$	&$2.29\times 10^{25}$	&$3.01\times 10^{25}$	&$2.29\times 10^{25}$	\\ \hline
Case 3 Red	&$2.32\times 10^{26}$	&$2.32\times 10^{26}$	&$2.30\times 10^{26}$	&$2.30\times 10^{26}$	\\
Case 3 Green	&$1.45\times 10^{25}$	&$1.45\times 10^{25}$	&$1.44\times 10^{25}$	&$1.44\times 10^{25}$	\\ \hline 
Case 4 Red	&$5.59\times 10^{25}$	&$8.36\times 10^{25}$	&$1.07\times 10^{26}$	&$4.86\times 10^{25}$	\\
Case 4 Green	&$1.09\times 10^{25}$	&$1.48\times 10^{25}$	&$1.73\times 10^{25}$	&$9.75\times 10^{24}$	\\ \hline \hline 
\end{tabular}
\end{table*} 

Case 0, representing the mass mechanism and displayed in the Figs. $1-4$ with a blue line, is the most studied mechanism in the literature. The value of the effective neutrino mass parameter $\left< \nu \right> =\left| \eta_{\nu} \right|$ is chosen to correspond to a neutrino mass limit of about 0.1 eV, which results in a calculated half-life of $9.4\times 10^{25}$, just beyond the current experimental limits but within the SuperNEMO reach \cite{nemo32014}. From Figs. $1-2$, one can see that this mode dominates the other contributions as long as $\left< \nu \right> \geq 4 \times \left< \lambda \right>$ and $\left< \nu \right> \geq 400 \times \left< \eta \right>$ (the red bands). 
Should any of the $\left< \lambda \right>$ or $\left< \eta \right>$ parameters increase four times (hatched green bands), the distributions change and one 
could identify the domination of another mechanism.

\begin{figure}[ht!]
{\includegraphics[width=0.98\linewidth]{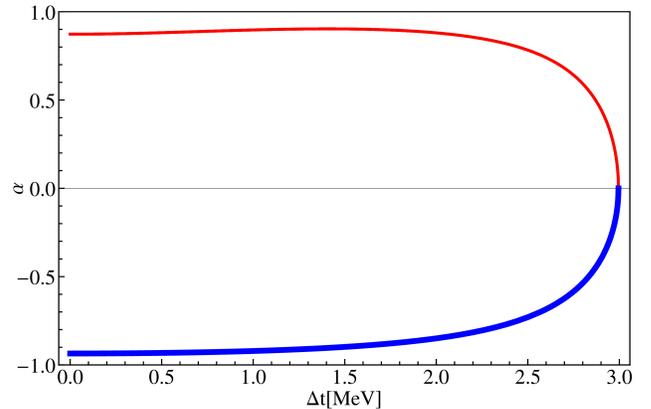}}
\caption{The angular correlation coefficient corresponding to Case 3. The meaning of the bands is the same as in Figs. $1-4$. The red and green (narrow) bands are overlapping.}
\end{figure}

\begin{figure}[ht!]
{\includegraphics[width=0.9\linewidth]{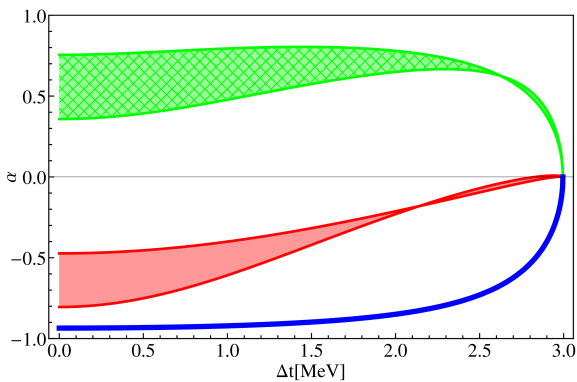}}
\caption{The angular correlation coefficient corresponding to Case 4. The meaning of the bands is the same as in Figs. $1-4$.}
\end{figure}

Case 1 presented in Fig. \ref{eta}  describes the $\eta$ mechanism dominance (hatched green bands) showing a significant change in the shape of the angular distribution (Fig. \ref{eta}, upper panel), while the energy distribution retains the shape of Case 0, only increasing in amplitude.
In the scenario of Case 2 presented in Fig. \ref{lambda}, one can see the dominance of the $\lambda$ mechanism (hatched green bands) in both distributions as changes in the shape and amplitude. One can conclude that one can use these different shape changes to distinguish between $\left< \nu \right>$, $\left< \lambda \right>$, and $\left< \eta \right>$ mechanism dominance, assuming that only two of them can compete. 

However, one needs to consider the case when the $|m_{ee}|=m_e \left< \nu \right>$ is very small or zero, while the $\lambda$ and $\eta$ mechanisms are competing. This scenario is covered by Case 3 presented in Fig. \ref{no_mass}. The interference term $C_{\lambda \eta}$ is very small leading to very narrow interference bands. 
Dominance of any of the two mechanisms would show little difference from the similar behavior shown in Figs. \ref{eta} and \ref{lambda} (the shape is fixed by the small interference term, while in Case 1 and 2 the dependence on the interference phases could distort the shapes).
The green lines in Case 3 are just rescaling of the red to emphasize the effect of rescaling relative to the standard mass mechanism (blue line). 
The shapes of the distributions and their changes seem to be similar to some of those in Fig. \ref{lambda}. However, the ratio max/min in the angular distribution (15/1 for Case 3 vs 2/1 for Case 2) could be used to distinguish between these two cases. 

Case 4 allows competition between all three contributions (Fig. \ref{all_mech}). Obviously, the qualitative behavior of these distributions cannot be easily disentangled from those of Cases $1-3$ discussed above. That would require a numerical simulation that includes interference effects to rule in or out some of these scenarios.

One should also mention that the energy distribution of the angular correlation coefficient, $\alpha = B(\epsilon)/A(\epsilon)$ in our Eq. (\ref{diff-rate}), could provide additional information (see, e.g., Figs. $6.5 - 6.9$ of \cite{Doi1985} and Fig. 7 of \cite{Stefanik2015}). Figures $5 - 8$ show the angular correlation coefficient $\alpha(\Delta t)$ of all four cases analyzed in Figs. $1 - 4$. One can clearly see that Cases 2 and 3 can also be identified by the value of $\alpha$ when the energies of the two emitted electrons are very close ($\Delta t \approx 0$). Cases 3 and 1 can be separated by the shape of their energy distributions. Figures $5 - 8$ show that the angular correlation coefficient could be also used to better identify the other cases analyzed in Figs. $1 - 4$.

Given the complexity of our analysis, and considering the potential usefulness for future analyses, we provide a link to a Mathematica file that can be used to perform these calculations and produce the plots included in this paper \cite{edistribMN2015}.


\begin{table*}[htb]
 \caption{Calculated limits of half-lives ratios, Eq. (\ref{hlratio}), for different combinations of isotopes (see text for details). For example, in the combination Ge/Se (1) corresponds to Ge and (2) to Se.}
 \begin{tabular}{@{}l|cc|cc|cc|cc|cc|cc@{}}  \hline \hline \label{hlrtable} 
 & \multicolumn{2}{c}{Ge/Se} & \multicolumn{2}{|c}{Ge/Te} & \multicolumn{2}{|c}{Ge/Xe} & \multicolumn{2}{|c}{Se/Te} & \multicolumn{2}{|c}{Se/Xe} & \multicolumn{2}{|c}{Te/Xe}  \\
\cline{2-13}
		& Ge & Se & Ge & Te & Ge & Xe & Se & Te & Se & Xe & Te & Xe \\ \hline
$G^{0\nu}_{01}\times10^{14}$ & 0.237 & 1.018 & 0.237 & 1.425 & 0.237 & 1.462 & 1.018 & 1.425 & 1.018 & 1.462 & 1.425 & 1.462 \\
$M^{0\nu}(1/2)$  & 3.57 & 3.39 & 3.57 & 1.93 & 3.57 & 1.76 & 3.39 & 1.93 & 3.39 & 1.76 & 1.93 & 1.76 \\
$M^{0N}(1/2)$  & 202 & 187 & 202 & 136 & 202 & 143 & 187 & 136 & 187 & 143 & 136 & 143 \\
\cline{2-13}
$T_{1/2}^{\nu}(1)/T_{1/2}^{\nu}(2)$ & \multicolumn{2}{c}{3.87} & \multicolumn{2}{|c}{1.76} & \multicolumn{2}{|c}{1.50} & \multicolumn{2}{|c}{0.45} & \multicolumn{2}{|c}{0.39} & \multicolumn{2}{|c}{0.85}  \\
$T_{1/2}^{N}(1)/T_{1/2}^{N}(2)$     & \multicolumn{2}{c}{3.68} & \multicolumn{2}{|c}{2.73} & \multicolumn{2}{|c}{3.09} & \multicolumn{2}{|c}{0.74} & \multicolumn{2}{|c}{0.84} & \multicolumn{2}{|c}{1.13}  \\
$R(N/\nu)$ present & \multicolumn{2}{c}{0.95} & \multicolumn{2}{|c}{1.55} & \multicolumn{2}{|c}{2.06} & \multicolumn{2}{|c}{1.63} & \multicolumn{2}{|c}{2.17} & \multicolumn{2}{|c}{1.33}  \\ \hline
$R(N/\nu)$ \cite{FaesslerGonzales2014} & \multicolumn{2}{c}{1.02} & \multicolumn{2}{|c}{1.39} & \multicolumn{2}{|c}{1.42} & \multicolumn{2}{|c}{1.36} & \multicolumn{2}{|c}{1.39} & \multicolumn{2}{|c}{1.03}  \\
\hline \hline
\end{tabular} 
\end{table*}

\section{Disentangling the heavy neutrino contribution}\label{heavy}

As mentioned in Sec. \ref{formalism}, 
if the $\eta_{\lambda}$ and $\eta_{\eta}$ contributions could be ruled out by the two-electron energy and angular distributions analyzed in the previous section, and in that case assuming a seesaw type I dominance \cite{DevMitra2015} the half-life is given by Eq. (\ref{hlifetime}).
Then, the relative contribution of the $\eta_{\nu}$ and $\eta^R_{N_R}$ terms can be identified if one measures the half-life of at least two isotopes \cite{Faessler2011,Vergados2012}, provided that the corresponding matrix elements $M^{0\nu}$ and $M^{0N}$ are known with good precision. 
References \cite{Faessler2011,Vergados2012} already provided some limits of the ratios of the half-lives  of different isotopes based on older QRPA calculations. However, based on those calculations, the two limits for $r(\nu/N)\equiv T_{1/2}^{\nu/N}(1)/T_{1/2}^{\nu/N}(2)$
\begin{equation}
\label{hlratio}
r(\nu/N)=\frac{G^{0\nu}_{01}(2) \left| M^{0\nu/N}(2) \right|^2}{G^{0\nu}_{01}(1) \left| M^{0\nu/N}(1) \right|^2},
\end{equation}
were too close to allow for a good separation of the contribution of these two mechanisms. In Eq. (\ref{hlratio}), terms $(1)$ and $(2)$ designate members of a pair of isotopes. Below, we present the results based on our shell model calculations given in Tables III and IV of Ref. \cite{NeacsuHoroi2015}. In Table \ref{hlrtable},  Ge, Se, Te, and Xe are short-hand notions for $^{76}$Ge, $^{82}$Se, $^{130}$Te, and $^{136}$Xe, respectively. In the table, we only use the NME calculated with CD-Bonn short-range correlations. The $G^{0\nu}_{01}$ factors from Table III of Ref. \cite{Stefanik2015} were used (they are very close to those of Ref. \cite{Mirea2014up}).

The pre-last line in Table \ref{hlrtable} presents the ratio of the ratios of half-lives, $R(N/\nu)=r(N)/r(\nu)$, calculated with our NME. One can see that the largest ratio is obtained for the combination $^{82}$Se/$^{136}$Xe. Its magnitude larger than 2 indicates that one can differentiate between these two limits if the half-lives are known with reasonable uncertainties and provided that the NME can be calculated with sufficient precision. The last line in Table \ref{hlrtable} shows the same quantity calculated with the recent QRPA NME taken from Table I (columns d) of Ref. \cite{FaesslerGonzales2014}. On can see that  these ratios are not as favorable in identifying the two limits. This analysis emphasizes again the need of having reliable NME for all mechanisms.

\section{Conclusions}
In this paper, we calculate nuclear matrix elements, phase-space factors, and half-lives for the  $\ 0\nu\beta\beta \left(0^+ \rightarrow 0^+ \right)$ decay of $^{82}$Se under different scenarios that include, besides the mass mechanism, the mixed right-handed/left-handed currents contributions known as $\eta$ and $\lambda$ mechanisms. For the mass mechanism dominance scenario, the results are consistent with previous calculations \cite{SenkovHoroiBrown2014} using the same Hamiltonian. Inclusion of contributions from $\eta$ and $\lambda$ mechanisms have the tendency to decrease the half-lives.

We present the two-electrons angular and energy distributions for five theoretical scenarios of mixing between mass mechanisms contributions and $\eta$ and $\lambda$ mechanism contributions. From the figures presented in the paper, one can recover the general conclusion \cite{Doi1985} that the energy distribution can be used to distinguish between the mass mechanism and the $\lambda$ mechanism, 
while the angular distribution can be used in addition to the energy distribution to distinguish between the mass mechanism and the $\eta$ mechanism, but the identification could be more nuanced due to the lack of knowledge of the interference phases.  
In the case of the energy distributions for the mass mechanism dominance (blue line) and the $\lambda$ mechanism dominance (green band in Figure \ref{lambda}, lower panel), we find similar results to those of Fig. 2 in Ref. \cite{snemo2010}. 
However, our results emphasize the significant role of the interference phases $\phi_1$ and $\phi_2$ in identifying the effect.

We also find out from the analysis of Case 3 that if the effective neutrino mass is very small, close to zero, and the $\eta$ and $\lambda$ mechanisms are competing, then one can potentially  identify this scenario from the $\lambda $ dominance, Case 2, by comparing the ratio min-to-max in the angular distributions and/or by the behavior of the angular correlation coefficient for almost equal electron energies. The small interference effects in Case 3 could be also used as an additional identification tool. These conclusions seem to be stable even if one considers small NME changes, such as those due to different short-range correlations  models.

We conclude that the $\eta$ mechanism, if it exists, may be favored to compete with the mass mechanisms due to the larger contribution from the phase-space factors. Reference \cite{Barry2013} shows however that it is possible to obtain a $\lambda$ mechanism dominance in some cases.  

Finally we show that if the $\eta_{\lambda}$ and $\eta_{\eta}$ contributions could be ruled out by the two-electron energy and angular distributions, the mass mechanisms can be disentangled from the heavy right-handed neutrino-exchange mechanism using ratios of half-lives of few isotopes. The analysis based on our shell model NME indicates that the most favorable combinations of isotopes would be $^{82}$Se/$^{136}$Xe and $^{76}$Ge/$^{136}$Xe.

Certainly, the analysis presented in this paper is based on the positive detection of the neutrinoless double-beta decay, followed by the collection of enough events that one can use to make assessments on the angular and energy distributions. Similar distributions were obtained with high precision by NEMO-3 for the $2\nu\beta\beta$ of $^{100}$Mo, but a very large number, about 1 million, of events were collected \cite{nemo32014}. Clearly, this large number of events will not be available for any $0\nu\beta\beta$ experiment, but we believe that the tools provided by our analysis could help to assess probabilities for these mechanisms even if only tens of events are collected.

\begin{acknowledgments}
Support from the NUCLEI SciDAC Collaboration under U.S. Department of Energy Grants No. DE-SC0008529 and DE-SC0008641 is acknowledged.
M.H. also acknowledges U.S. NSF Grant No. PHY-1404442.
\end{acknowledgments}

\appendix

\section{Left-right symmetric model} \label{app:cp}

Left-right symmetric models \cite{PatiSalam1974,MohapatraPati1975,Senjanovic1975,KeungSenjanovic1983} could explain the physics of the 
right-handed currents, which may contribute to the neutrinoless double-beta decay process, and are also under current investigation at LHC \cite{CMS2014}. Specific details for double-beta decay can be found in Ref. \cite{Barry2013}.

The neutrino mixing matrices 
are defined by
\begin{equation}
 n\sp{\prime}_L=\binom{\nu\sp{\prime}_L}{{\nu\sp{\prime}_R}^c}=\binom{U \ S}{T \ V}\binom{\nu_L}{N_R^c},
\end{equation}
where $\nu\sp{\prime}_L$, ${\nu\sp{\prime}_R}^c$ are flavor eigenstates, and $\nu_L$, $N_R^c$ are mass eigenstates. Here, the $U$ and $V$ matrices are almost unitary, while the $S$ and $T$ matrices are very small. The sterile neutrinos $\nu\sp{\prime}_R$ and the mass eigenstates $N_R$ are presumed to be very heavy, but at least the lightest ones are at the TeV scale. Light (1 eV) sterile neutrinos could exist, and they could influence the effective neutrino mass and the outcome of $0\nu\beta\beta$ decay \cite{Vergados2012}, but they may be detected in neutrino oscillations experiments.
The neutrino physics parameter $\left| \left< m_{ee} \right> \right| \equiv \left| \sum U_{ek}^2 m_k \right|$ is the effective  electron neutrino mass, and the suitably
normalized dimensionless parameter that describes lepton number violation is (the upper limits for the neutrino physics parameters below were taken from Refs. \cite{Stefanik2015,Barry2013})
\begin{equation}
 \left| \eta_\nu \right| = \frac{\left| \left< m_{ee} \right> \right| }{m_e}=\frac{\left| \sum_k^{light} U_{ek}^2 m_k \right|}{m_e} \lesssim 7 \times 10^{-7},
\end{equation}
with $U_{ei}$ the (PMNS) mixing matrix of light neutrinos, $m_i$ the light neutrino masses, and $m_e$ the electron mass. For the mixing of the left- and right-handed currents with the heavy neutrino the neutrino physics parameters in the left-right symmetric model are given by
\begin{equation}
 \left| \eta^L_{N_R} \right| = m_p  \left| \sum_k^{heavy} \frac{{S_{ek}}^2}{M_k} \right| \lesssim 7 \times 10^{-9},
\end{equation}
\begin{equation}
 \left| \eta^R_{N_R} \right| = m_p \left( \frac{m_{W_L}}{m_{W_R}} \right)^4 \left| \sum_k^{heavy} \frac{{V^\star_{ek}}^2}{M_k} \right| \lesssim 7 \times 10^{-9},
\end{equation}
where $m_{W_R} \ (m_{W_L})$ is the mass of the right-handed $W_R \ (\textmd{left-handed}\ W_L)$, $M_i$ are the masses of the heavy neutrinos, and $V$ is
the right-handed analogue of the PMNS matrix U. To satisfy the present limit of $\left| \eta^R_{N_R} \right|$, one needs $m_{W_R}$ and some of the $M_k$ masses at TeV scale. For the terms that could contribute to the neutrinoless double-beta decay that involve a mixture of left-handed and right-handed currents, the $\eta_\lambda$ and $\eta_\eta$ neutrino physics parameters are

\begin{equation}
 \left| \eta_\lambda \right| = \left( \frac{m_{W_L}}{m_{W_R}} \right)^2 \left| \sum_k^{light} U_{ek}T^\star_{ek}\right| \lesssim 4 \times 10^{-7}\ ,
\end{equation}


\begin{equation}
 \left| \eta_\eta \right| = \textmd{tan}\xi \left| \sum_k^{light} U_{ei}T^\star_{ek} \right| \lesssim 3 \times 10^{-9}\ .
\end{equation}
The heavy neutrino contributions to both $\lambda$ and $\eta$ mechanisms are suppressed, being proportional to $\sum_k^{heavy} S_{ek}V^\star_{ek} q/ M^2_k$.

The $CP$ phases used in Eq. (\ref{lifetime}) are
\begin{eqnarray} \label{phases}
 \nonumber \phi_1 &=& \text{arg}\left[ \left(\sum_k^{light}U_{ek}^2 m_k \right) \left(\left( \frac{M_{W_L}}{M_{W_R}}\right)^2 \sum_k^{light}U_{ek}V_{ek} \right)^\star \right], \\
 \phi_2 &=& \text{arg}\left[ \left(\sum_k^{light}U_{ek}^2 m_k \right) \left( \xi \sum_k^{light}U_{ek}V_{ek} \right)^\star \right].
 \end{eqnarray}

\section{$0\nu\beta\beta$ NME} \label{app:appendixa}
Most of the theoretical formalism used in this work is adopted from Refs. \cite{Doi1985} and \cite{SuhonenCivitarese1998}, with little change of notation for 
simplicity and consistency wherever need.

The  $C_{1-6}$ factors composed from PSF and NME \cite{Doi1985} are
\begin{subequations} \label{ci-uri}
\begin{align}
C_1 &= \left( 1-\chi_F \right)^2 G_{1}, \\
C_2 &= -\left( 1-\chi_F \right)\left[ \chi_{2-} G_{3}-\chi_{1+} G_{4} \right], \\
\nonumber C_3 &= \left( 1-\chi_F \right) \\ 
    &\times \left[ \chi_{2+} G_{3}-\chi_{1-} G_{4} - \chi_P G_{5} + \chi_R G_{6} \right], \\
C_4 &= \left[ \chi_{2-}^2 G_{2}+\frac{1}{9}\chi_{1+}^2 G_{4}-\frac{2}{9}\chi_{1+}\chi_{2-}G_3 \right], \\
\nonumber C_5 &= \chi_{2+}^2 G_{2}+\frac{1}{9}\chi_{1-}^2 G_{4}-\frac{2}{9}\chi_{1-}\chi_{2+}G_3 +\chi_P^2 G_8\\ 
    &- \chi_P \chi_R G_7 + \chi_R^2 G_9 , \\
\nonumber C_6 &= -2\left[ \chi_{2-}\chi_{2+} G_{2}-\frac{1}{9}(\chi_{1+}\chi_{2+}+\chi_{2-}\chi_{1-}) G_{3}  \right. \\
    &+ \left. \frac{1}{9}\chi_{1+}\chi_{1-}G_4 \right], 
\end{align}
\end{subequations}

\begin{subequations}\label{chi}
\begin{align}
 \label{chi1} \chi_{1\pm}&=\chi_{GTq}\pm 3\chi_{Fq} - 6\chi_T, \\
 \label{chi2} \chi_{2\pm}&=\chi_{GT\omega}\pm \chi_{F\omega} - \frac{1}{9}\chi_{1\pm}.
\end{align}
 \end{subequations}

The normalized NME
\begin{equation}\label{chi-form}
 \chi_\alpha=M_\alpha/M_{GT}^{0\nu},
\end{equation}
where $\alpha=F,T,GT\omega,F\omega,GTq,Fq,R$, and $P$. All Fermi-type matrix elements $M_{F(\omega q)}$ are multiplied by $g_V/g_A$. 

Due to the two-body nature of the transition operator, the matrix elements are reduced to sums of products of two-body transition densities (TBTD)  and matrix elements for two-particle states \cite{HoroiStoica2010}:
\begin{eqnarray}\label{nme-gt}
\nonumber M_\alpha^{0\nu} & = & \sum_{j_p j_{p^\prime} j_n j_{n^\prime} J_\pi} TBTD \left( j_p j_{p^\prime} , j_n j_{n^\prime} ; J^\pi \right) \\
&\times & \left< j_p j_{p^\prime}; J^\pi \| \tau_{-1} \tau_{-2}O^\alpha_{12} \| j_n j_{n^\prime} ; J^\pi \right> .
\end{eqnarray}
The detailed expressions for the two-body transition operators $(O^\alpha_{12})$ can be found in Ref. \cite{Muto1989}. 
They can be factorized into products of coupling constants and operators which act on the intrinsic spin, relative and center-of-mass wave functions of 
two-particle states \cite{HoroiStoica2010}. 

The NME depend on four dimensionless neutrino potentials defined by the integral over the momentum of the virtual neutrino. 
Expressions for the Gamow-Teller (GT), the Fermi (F), and the tensor (T) cases are described in detail in Refs. \cite{HoroiStoica2010,SenkovHoroi2013}. 
The other three potentials are presented here in a form similar to Eq. (12) of Ref. \cite{Muto1989},
\begin{subequations}\label{potentials}
\begin{align}
&H_\omega^{0\nu}(r)	=\frac{2R}{\pi}\int_0^\infty \frac{ q^2 j_0(qr) \textmd{d}q}{\left( q+\left< E \right> \right)^2} \equiv \int_0^\infty q^2 j_0(qr) V_\omega(q) \textmd{d}q, \\
&\nonumber \textmd{for the }M^{0\nu}_{GTw}\textmd{ and }M^{0\nu}_{Fw}\textmd{ NME,}  \\
&H_q^{0\nu}(r)		=\frac{2R}{\pi}\int_0^\infty \frac{q^2 j_1(qr) \textmd{d}q}{ q+\left< E \right>} \equiv  \int_0^\infty q^2 j_1(qr) V_q(q) \textmd{d}q, \\
&\nonumber \textmd{for the }M^{0\nu}_{GTq}, M^{0\nu}_{Fq}, M^{0\nu}_{T},\textmd{ and }M^{0\nu}_{P}\textmd{  NME}. \\ 
&\nonumber \textmd{In the case of }M^{0\nu}_{R},\textmd{ the potential is written as:}\\
&H_R^{0\nu}(r)		=\frac{2R^2}{\pi M}\int_0^\infty \frac{q^3 j_0(qr)\textmd{d}q}{ q+\left< E \right>}\equiv \int_0^\infty q^2 j_0(qr) V_\omega(q) \textmd{d}q, 
\end{align}
\end{subequations}
where $M$ is the nucleon mass, $R$ the nuclear radius ($R=1.2A^{1/3} fm$), $\left< E \right>$ represents the closure energy, $V_{\omega,q,R}$ are the Fourier transforms of the potentials, and $j_\kappa(qr)$ are spherical Bessel functions of rank $\kappa$. Due to the small contribution of the $\chi_{P}$ term, we take a typical value of 0.5 for the associated normalized NME.

The computation of the matrix element requires solving a double integral over the coordinate space and over 
the momentum [from Eq. (\ref{potentials})] of the form \cite{NeacsuStoicaHoroi2012}
\begin{eqnarray} \label{i-alpha}
&& \nonumber \mathcal{I}_\alpha(\mu;m)  = \int_0 ^\infty q^2 dq \ V_\alpha(q)\\
&&\times \left( \frac{2}{\pi} \right)^{\frac{1}{2}} \left( 2 \nu \right) ^{\frac{m+1}{2}} \int_0 ^\infty dr \ e^{-\mu r^2}r^m j_\kappa(qr)
\end{eqnarray}
where $\mu$ = $\nu$, $\nu+a$, $\nu+2a$, with $\nu$ the oscillator constant and $m$ is an integer.

It was previously observed in Ref. \cite{Doi1985} that the three potentials in Eq. (\ref{potentials}) are formally divergent but the associated radial matrix elements are not, if certain  precautions are taken, such as first performing the radial integrals and then the integrals of the momentum in Eq. (\ref{i-alpha}), as was done in Ref. \cite{HoroiStoica2010}.

In Ref. \cite{SenkovHoroiBrown2014}, a method was proposed for obtaining an optimal closure energy, which yields similar results as when preforming calculations beyond the closure approximation. Here, we use an optimal average closure energy $\left< E \right>$ of $3.4$ MeV, which has been shown to produce accurate results in the case of 
$M_{GT}$ and $M_F$ (see Fig. 5 of Ref. \cite{SenkovHoroiBrown2014}). Therefore, our NME do not have any significant uncertainties related to choice of the closure energy. 
Higher order corrections of the nuclear current for the Gamow Teller nuclear matrix element and CD-Bonn parametrization short-range correlations are taken
into account as described in Ref. \cite{HoroiStoica2010}.

To calculate the two-electron angular and relative energy distributions, we take into account the decay rate as described by Eq. (C$\cdot $3$\cdot $1) of Ref. \cite{Doi1985}.
This leads to the expressions of Eqs. (\ref{diff-rate}) and (\ref{ab}). The factors $N_{1-4}(\epsilon_1)$ represent mixtures of NME and PSF, expressed as
\begin{subequations} \label{n-uri}
\begin{align}
N_1(\epsilon_1)&=a_{\textmd{-}1\textmd{-}1}^*\left[ \left( Z_1 - \frac{4Z_6}{3} \right) - \left( \frac{4}{m_eR}\right) \left(Z_4-\xi\frac{Z_6}{6} \right) \right], \\
N_2(\epsilon_1)&=a_{11}^*\left[ \left( Z_1 - \frac{4Z_6}{3} \right) + \left( \frac{4}{m_eR}\right) \left(Z_4-\xi\frac{Z_6}{6} \right) \right], \\
N_3(\epsilon_1)&=a_{1-1}^*\left[ \left( Z_1 - \frac{2Z_5 }{3}\right) - \left( \frac{\epsilon_{12}}{m_e}\right) \left(Z_3+\frac{Z_5 }{3} \right) \right], \\
N_4(\epsilon_1)&=a_{-11}^*\left[ \left( Z_1 - \frac{2Z_5 }{3}\right) + \left( \frac{\epsilon_{12}}{m_e}\right) \left(Z_3+\frac{Z_5 }{3} \right) \right],
\end{align}
\end{subequations}
with $ \xi=3\alpha Z_s+(T+2)m_e R$, $\epsilon_{12}=\epsilon_1-\epsilon_2$ and $ a_{\kappa\lambda}=\tilde{A}_\kappa(\epsilon_1)\tilde{A}_\lambda(\epsilon_2)$, where $\epsilon_2 = T+2-\epsilon_1.$

\begin{equation}
 \tilde{A}_{\pm k}(\epsilon)\cong\sqrt{(\epsilon \mp m_e)/2\epsilon} \sqrt{F_{k-1}(Z_s,\epsilon)}.
\end{equation}

\begin{eqnarray} \label{ffunction}
\nonumber F_{k-1}(Z_s,\epsilon)&=&\left[ \frac{\Gamma(2k+1)}{\Gamma(k)\Gamma(2\gamma_k+1)}\right] ^2 \\ 
&\times & (2pR)^{2(\gamma_k-k)}|\Gamma(\gamma_k+iy)|^2 e^{\pi y}
\end{eqnarray}

\begin{equation}
\gamma_k=\sqrt{k^2-(\alpha Z)^2}, \qquad y=\alpha Z_s \epsilon /p,
\end{equation}
where $\alpha$ is the fine structure constant, $p_i=\sqrt{\epsilon_i^2-1} ($with $i=1,2)$, and $Z_s= Z \cdot \frac{S_f}{100}$ the ''screened`` charge of the final nucleus, 
and $S_f=92$ is the effective ''screening`` factor from Table 4 of Ref. \cite{HoroiNeacsuPSF2015}.
Here, $Z_{1-6}$ are composed of the NME from Eq. (\ref{nme-gt}), defined as follows:
\begin{subequations}
\begin{align}
Z_1&=(\left< \nu\right>) (\chi_F-1)M_{GT}^{0\nu},\\
Z_3&= \left[-\left< \lambda \right> (\chi_{GT\omega}-\chi_{F\omega})e^{-i\phi_1} \right. \\ 
&+ \left. \left<\eta\right>(\chi_{GT\omega}+\chi_{F\omega})e^{-i\phi_2} \right] M_{GT}^{0\nu},\\
Z_4&=\left<\eta\right>\chi_R e^{-i\phi_2} M_{GT}^{0\nu},\\
Z_5&= \frac{1}{3}\left[ \left< \lambda \right> \chi_{1+}e^{-i\phi_1} - \left< \eta \right> \chi_{1-}e^{-i\phi_2} \right] M_{GT}^{0\nu},\\
Z_6&=\left< \eta \right>\chi_P e^{-i\phi_2}M_{GT}^{0\nu}.
\end{align}
\end{subequations}

\section{$0\nu\beta\beta$ decay PSF expressions} \label{app:appendixb}
The PSF are calculated using the following expression adopted from Eq. (A.27) of Ref. \cite{SuhonenCivitarese1998}:
\begin{flalign} \label{psfs-form}
 G_k=\frac{a_{0\nu}}{\textmd{ln}2(m_e R)^2}\int\limits_1^{T+1} b_k F_0(Z_s,\epsilon_1)F_0(Z_s,\epsilon_2)\omega_{0\nu}(\epsilon_1) \text{d}\epsilon_1, & &
\end{flalign}
where $R$ is the nuclear radius ($R=r_0 A^{1/3}$, with $r_0=1.2$ fm) and $F_0$ is defined in Eq. (\ref{ffunction}) for $k=1$.
\begin{equation} \label{a0}
a_{0\nu}=\frac{g_A^4\left( G_F \textmd{cos}\theta_c \right) ^4 m_e^9}{32\pi^5},
\end{equation}
with $G_F=1.1663787\times 10^{-5}\ GeV^{-2}$ the Fermi constant, and $\textmd{cos}\theta_c=0.9749$ the Cabbibo angle.
In Ref. \cite{SuhonenCivitarese1998}, the constant $g^{0\nu}=a_{0\nu}/\textmd{ln}2=2.8 \times 10^{-22} g_A^4\ yr^{-1}$ was used.
Taking into account the value $g_A=1.27$, instead of $g_A=1.254$, would change the results by $5\%$. One should mention that the $G^{0\nu}_{01} g^4_A$ product in Eq. \ref{hlifetime} is equal to $G_1$.  Also, in Eq. (C1)

\begin{equation} \label{omega0} 
\omega_{0\nu}(\epsilon_1)= p_1 p_2 \epsilon_1 \epsilon_2,
\end{equation}
with $ \epsilon_2=T+2-\epsilon_1$, $p_{1,2}=\sqrt{\epsilon_{1,2}^2-1}$, and $T$ defined in Eq. (\ref{t-energy}).

The kinematical factors $b_k$ are defined as
\begin{flalign}
\ \ \ \ b_1&=1,&
\end{flalign}

\begin{flalign}
\ \ \ \ b_2&=\frac{1}{2}\left( \frac{\epsilon_1 \epsilon_2 -1}{\epsilon_1 \epsilon_2}\right) (\epsilon_1-\epsilon_2)^2, &
\end{flalign}

\begin{flalign}
\ \ \ \ b_3&= (\epsilon_1-\epsilon_2)^2 / \epsilon_1 \epsilon_2 ,&
\end{flalign}

\begin{flalign}
\ \ \ \ b_4&= \frac{2}{9} \left( \frac{\epsilon_1 \epsilon_2 -1}{\epsilon_1 \epsilon_2}\right) , &
\end{flalign}

\begin{flalign}
\ \ \ \ b_5&= \frac{4}{3} \left(\frac{(T+2)\xi}{2r_A\epsilon_1\epsilon_2} -\frac{\epsilon_1 \epsilon_2 +1}{\epsilon_1 \epsilon_2}\right), &
\end{flalign}

\begin{flalign}
\ \ \ \ b_6&= \frac{4(T+2)}{r_A\epsilon_1\epsilon_2} ,&
\end{flalign}

\begin{flalign}
\ \ \ \ b_7&= \frac{16}{3} \frac{1}{r_A\epsilon_1\epsilon_2}\left( \frac{\epsilon_1\epsilon_2+1}{2r_A} \xi-T-2 \right) ,&
\end{flalign}

\begin{flalign}
\ \ \ \ \nonumber b_8&= \frac{2}{9} \frac{1}{(r_A)^2\epsilon_1\epsilon_2}\left[(\epsilon_1\epsilon_2+1)(\xi^2+4(r_A)^2) \right. & \\
   & \left. - 4r_A \xi(T+2) \right] , &
\end{flalign}

\begin{flalign}
\ \ \ \ b_9&= \frac{8}{(r_A)^2}\left( \frac{\epsilon_1\epsilon_2+1}{\epsilon_1\epsilon_2} \right) ,& 
\end{flalign}
with $\xi=3\alpha Z_s+r_A(T+2)$, where $\alpha$ represents the fine structure constant, $Z_s= Z \cdot \frac{S_f}{100}$ the "screened" charge of the final nucleus, and $r_A=m_e R$.

In Eq. (\ref{flifetime}), Eq. (\ref{hlifetime}), Eq. (\ref{hlratio}) and in Table \ref{hlrtable} we use the factor $G_{01}^{0\nu} = G_1/\left( g_A \right)^4$.

\bibliographystyle{apsrev}
\bibliography{bbmech}

\end{document}